\documentclass[conference]{IEEEtran}
\IEEEoverridecommandlockouts
\usepackage{cite}
\usepackage{amsmath,amssymb,amsfonts}
\usepackage{algorithmic}
\usepackage{graphicx}
\usepackage{textcomp}
\usepackage{xcolor}
\usepackage[colorlinks=true,urlcolor=teal,
            linkcolor=teal,citecolor=teal]{hyperref}
\usepackage{url}

\usepackage{braket}
\usepackage{textcomp}
\usepackage{bm}
\usepackage{csquotes}
\usepackage{censor}
\usepackage{etoolbox}
\usepackage{xspace}
\usepackage{xstring}
\usepackage{algorithm}
\usepackage{algorithmic}

\newcommand{\eg}{\emph{e.g.}\xspace}

\newcommand{\etal}{\emph{et al.}\xspace}

\newif\ifdraft
\drafttrue
\ifdraft
\newcommand{\kwnote}[1]{ {\textcolor{green} { ***Karen: #1 }}}

\else
\newcommand{\note}[1]{}
\newcommand{\omnote}[1]{}
\newcommand{\kwnote}[1]{}
\fi

\newboolean{anonymous}
\setboolean{anonymous}{false}
\ifbool{anonymous}{}{}
\ifbool{anonymous}{}{}
\ifbool{anonymous}{}{}

\begin{document}

\title{A Reality Check on Quantum Optimisation: Evidence from an Industrial Case Study}


\author{
\IEEEauthorblockN{
Hila Safi\IEEEauthorrefmark{1}\IEEEauthorrefmark{2}
Karen Wintersperger\IEEEauthorrefmark{1},
Oliver von Sicard\IEEEauthorrefmark{2},
Christoph Niedermeier\IEEEauthorrefmark{1},
Wolfgang Mauerer\IEEEauthorrefmark{2}\IEEEauthorrefmark{1}
}

\IEEEauthorblockA{\IEEEauthorrefmark{1}
Siemens Foundational Technologies, Munich, Germany\\
\{hila.safi, karen.wintersperger, christoph.niedermeier\}@siemens.com
}

\IEEEauthorblockA{\IEEEauthorrefmark{2}
Technical University of Applied Sciences Regensburg, Laboratory for Digitalisation, Regensburg, Germany\\
\{oliver.vonsicard, wolfgang.mauerer\}@oth-regensburg.de
}

\thanks{\textsuperscript{*} Oliver von Sicard conducted this work while employed at Siemens AG.}
}


\maketitle

\begin{abstract}
Quantum Processing Units promise speed-ups for selected computational
problems, including combinatorial optimisation, but their industrial utility
remains an open challenge. We study an industrial variant of the Job-Shop
Scheduling Problem using quantum, quantum-inspired, and classical methods
across three platforms: IBM Quantum, the D-Wave Quantum Annealer, and the
Fujitsu Digital Annealer. By tailoring formulations to hardware-specific
constraints, we show that hardware-software co-design is essential for solution
quality and scalability. We benchmark all approaches against an exact classical
solver and a MILP formulation, evaluating runtime, solution quality, and
scalability. Our results indicate that quantum and quantum-inspired optimisation
can support industrial solver selection, integration in classical
workflows, modelling decisions, and early
proof-of-concept development, while suggesting a potential path towards improved
approximations for industrial scheduling.
\end{abstract}

\section{Introduction}
Quantum computers may solve selected computational problems exponentially faster
than classical systems, but their practical value can only be assessed
on real-world applications~\cite{bayerstadler:2021,Yue:2023}. Concurrently, quantum engineering increasingly highlights the need to consider the full computational stack, including algorithm design,
problem formulation, hardware constraints, and classical post-processing. 
In this work, we study how these factors affect solution quality~\cite{Gabor:2019}
and how quantum and quantum-inspired methods can be integrated
into existing classical industrial workflows. Our comparison is application-driven,
we focus on a real industrial variant
of the Job Shop Scheduling Problem (JSSP), a central optimisation challenge in
production logistics and manufacturing, where efficiency and cost reduction are
key objectives~\cite{Arisha2001JobSS}. The industrial instances provided
by Siemens AG allow us to evaluate how problem encodings, solver paradigms,
and hardware characteristics influence end-to-end performances and affect
practical scheduling solutions.
We use this case study to derive transferable design considerations for near-term
optimisation, particularly regarding the trade-off between fidelity,
constraint density, and hardware tractability, while also contributing
to a better understanding of how quantum computing may be applied to
real-world optimisation problems as hardware matures.
The study therefore serves not only as
a benchmark of current performance, but also as a first step towards
development and the preparation of future industrial optimisation pipelines.
As reference points, we develop a classical exact solver for small tractable
instances and a Mixed-Integer Linear Programming (MILP) model for larger ones.
While MILP is exact in principle, typically using branch-and-bound and
branch-and-cut~\cite{Mitchell2025}, we solve larger instances under a fixed time
budget and use the best solution found as an approximate classical
baseline~\cite{gurobi2026methods}. This provides a scalable and practically
relevant comparator for quantum-inspired optimisation.
We compare two QUBO formulations for the JSSP. The Single-Constraint Model uses
a compact constraint structure, while the Multi-Constraint Model introduces
additional constraints to represent scheduling dependencies more accurately.
This exposes the trade-off between modelling simplicity, constraint density, and
hardware tractability. The models are evaluated on three platforms: IBM
gate-based quantum hardware, the D-Wave Advantage quantum annealer, and the
Fujitsu Digital Annealer. Our results show that the number of constraints and
the degree of hardware proximity in the formulation strongly affect runtime and
solution quality across all platforms.
More broadly, the findings show that hardware performance cannot be separated from
modelling choices and device characteristics. This underscores the importance of
hardware-software co-design and systematic design-space exploration across the
full quantum stack, from model formulation to device-level execution. 
This paper is augmented by a \href{https://github.com/lfd/reality_check_qce_26.git}{reproduction package}~\cite{Mauerer:2022},
which is permanently archived on Zenodo at
\href{https://doi.org/10.5281/zenodo.21363322}{doi:10.5281/zenodo.21363322}
(links available in the PDF).

\section{Related Work}
The Job Shop Scheduling Problem (JSSP) is a fundamental problem in
production logistics and operations research and is well known to be
computationally hard, even restricted variants are NP-hard, and
realistic instances combine machine-capacity constraints,
sequence-dependent setup times, and objective
trade-offs~\cite{Arisha2001JobSS,jssp_complexity}. Classical research on
JSSP and related parallel-machine scheduling has produced a large body
of exact and heuristic methods, including branch-and-bound and MILP
formulations. For parallel machines with sequence-dependent setup times
(PMSP-SDST), Ying~\etal\ provide a recent survey of modelling and
algorithmic strategies~\cite{YING2025100340}. Our Classical
Approximation Model follows this PMSP-SDST literature and uses a
time-limited MILP formulation to obtain practically comparable
approximations, similar in spirit to the approach of
Schönberger~\etal\ for join-order
optimisation~\cite{schoenberger:25:pvldb}.
On the quantum side, QAOA was introduced by Farhi~\etal\ as a
variational hybrid algorithm for combinatorial optimisation on NISQ
devices~\cite{Fahri_2014}, and has received intensive scrutiny since (see Refs.~\cite{blekos_review_2024,zhou2020quantum} and references therein); modifications to the structure of the quantum 
circuit itself (\eg, Refs.~\cite{Zhang:2017,Wang:2020,Baertschi:2020,
bravyi2020obstacles,zhou2020quantum}) and changes to the classical 
optimisation procedure (\eg, 
Refs.~\cite{Egger:2021,Awasthi:2023,Tate:2023,Vijendran:2024,Sud:2024,
montanez2024towards,Fingar:2024,streif2020training,kruger:2025}) have been proposed to improve
performance especially in NISQ scenarios.
QAOA and related methods have also
been studied for countless graph and scheduling problems, including evaluations on 
industrial data and the effects of embedding choices~\cite{schmidbauer:25:qce}. Our work complements this
line by studying QAOA-based JSSP formulations using a systematic evaluation approach~\cite{Lorenz:2025,Gierisch:2026} on IBM hardware.
Quantum annealing and quantum-inspired optimisation have also been
explored for combinatorial optimisation problems. D-Wave annealers have
been applied to scheduling and logistics via Ising and QUBO encodings
that must be embedded into sparse hardware
graphs~\cite{johnson2011quantum,boothby2020next,Krueger:2020} (and are known to possibly benefit from changes in the problem formulation itself~\cite{Schmidbauer:2025,Schmidbauer:2026,Sax:2020}), while Fujitsu’s
Digital Annealer has been proposed as a quantum-inspired architecture
for dense QUBO problems~\cite{aramon2019physics,fujitsuDAwhitepaper}.
Our work extends this literature by comparing the Digital Annealer not
only with classical MILP and heuristic baselines, but also with a
physical quantum annealer (D-Wave) and gate-based quantum devices
(IBM), using the same industrial JSSP instances.
Scheduling-oriented quantum optimisation has been studied in both
annealing and gate-based settings. Venturelli~\etal\ introduced an
early D-Wave QUBO formulation for
JSSP~\cite{venturelli2016quantumannealingimplementationjobshop},
Schworm~\etal\ extended annealing approaches to flexible job-shop
scheduling in manufacturing~\cite{SchwormWuGlattetal.2024},
Orts~\etal\ considered the unrelated parallel machine
setting~\cite{Orts_2022}, Kurowski~\etal\ investigated QAOA for
JSSP~\cite{Kurowski_23}, and Schwenzow~\etal\ studied VQE for identical
parallel machines with sequence-dependent setup
times~\cite{Schwenzow2024}. Together with earlier work on quantum
scheduling by Lu~\etal~\cite{Feng07} and annealing-based load balancing
by Rathore~\etal~\cite{Rathore_25}, these studies show that formulation
choices and hardware constraints are central to quantum optimisation for
scheduling.
Finally, our study is closely related to work on hardware-software
co-design and quantum scalability~\cite{safi2023}.

\section{Problem Description and Models}
This paper considers a JSSP variant based on a real-world industrial
Siemens application. We consider \(N\) jobs, each executed once, and
\(M\) identical production machines on which any job can run. Jobs differ
in processing time and require a specific tool configuration, referred to
as a rig. We consider \(L\) possible rigs:
\begin{itemize}
    \item Jobs: \( i \in \{1, 2, \ldots, N\} \)
    \item Machines: \( j \in \{1, 2, \ldots, M\} \)
    \item Rigs: \( k \in \{1, 2, \ldots, L\} \)
\end{itemize}
Some jobs require the same rig, while others require different rigs.
Since rig changes take time and depend on the current and subsequent job,
jobs with the same rig should preferably be processed consecutively on
the same machine. Each machine is assumed to start with an initial rig
from prior production. The objective is to minimise the makespan, defined
as the maximum production time across all machines.
Modelling the complete scheduling task as a QUBO requires on the order
of $N^2M$ variables, as the time domain must also be represented.
Moreover, rig-change optimisation is difficult to encode because it
requires identifying whether two jobs are executed consecutively on the
same machine. With $N$ global time steps, consecutive jobs on one
machine do not necessarily correspond to indices $t$ and $t+1$.
We therefore split the task into two stages. First, jobs are assigned to
machines based on durations and rigs using a quantum or classical solver.
Second, the jobs on each machine are ordered to minimise rig changes;
for the considered problem sizes, this step can be solved classically
and is not studied further.
Since the maximum function cannot be directly encoded in a QUBO, we
approximate makespan minimisation by balancing total production times
across machines. We evaluate two QUBO assignment models, a
Single-Constraint and a Multi-Constraint formulation, which differ in
objective design and constraint density. Because the task is split, an
optimal QUBO solution does not necessarily recover the global optimum of
the full scheduling problem. A classical MILP-based Classical
Approximation Model is introduced as an additional baseline.
\subsection{Single-Constraint Model}\label{subsec:model-1} 
The first model is a formulation with one constraint.
We minimise the difference of the overall duration on each machine
to the mean production duration $D$ defined as
\begin{equation}
    \bar{D} = \frac{1}{M}\sum_{i=1}^Nd_i,
\end{equation}
where $d_i$ is the duration of job $i$.
The total duration on each machine includes job durations and rig-change times,
including the change from the machine's initial rig. Since the job order is not
optimised, the first job is unknown; we therefore approximate this initial setup
cost using the average rig-change time between all assigned jobs and the initial
rig.
Our binary decision variable is
\begin{align}
x_{ij} &=
\begin{cases}
1, & \text{if job } i \text{ is processed on machine } j,\\
0, & \text{otherwise}.
\end{cases}
\label{eq:xij}
\end{align}
The matrix $R \in \mathbb{R}^{N \times N}$ contains the rig-change times 
between pairs of jobs. The matrix $S \in \mathbb{R}^{N \times M}$ describes
the rig-change time difference between each job and the initial rig of each machine.
The average rig-change time for machine $j$ is
\begin{equation}
    S_j := \frac{1}{N} \sum_{i=1}^N S_{ij}.
\end{equation}
Objective 1 $|$ $H_{\mathrm{obj}}$: Minimise the difference to
the mean duration over all machines, including the initial rig-change time
\begin{equation}
 H_{\mathrm{obj}} = \sum_{j=1}^M \left(\sum_{i=1}^N x_{ij} (d_i +  S_{ij}) - (\bar{D}+S_j) \right)^2.
\end{equation}
Objective 2 $|$ $H_{\text{rig}}$: Minimise rig change time between
jobs and between the jobs and the initial rig on each machine
\begin{equation}
    H_{\text{rig}} = \sum_{j=1}^M \left(\sum_{i < i'}^N x_{ij} x_{i'j} R_{ii'}\right) + \left(\sum_{i=1}^N x_{ij} S_{ij}\right).
\end{equation}
Constraint $|$ $H_{\mathrm{single}}$: Each job must be assigned to exactly one
machine
\begin{equation}
    H_{\mathrm{single}} = \sum_{i=1}^N \left( \sum_{j=1}^M x_{ij} -1 \right)^2.
\end{equation}
\subsection{Multi-Constraint Model}\label{subsec:model-2} 
The Multi-Constraint Model extends the assignment problem of the 
Single-Constraint Model by explicitly introducing
rig configurations and separating the decision of
\begin{enumerate}
    \item which machine executes each job, and
    \item which rig configuration is active on that machine.
\end{enumerate}
To this end, we additionally consider the binary decision variable:
\begin{align}
y_{kj} &=
\begin{cases}
1, & \text{if rig configuration } k \text{ is needed on machine } j,\\
0, & \text{otherwise}.
\end{cases}
\label{eq:model14_y}
\end{align}
Furthermore, we define the following model parameters:
Let \(R_{\ell k}\) be the rig–change time from configuration \(\ell\) to \(k\) with \(R_{kk} = 0\).
The mean rig–change time associated with rig \(k\) is
\begin{equation}
  \bar R_k = \frac{1}{L-1} \sum_{\ell=1}^L R_{\ell k}.
  \label{eq:model14_Rbar}
\end{equation}
Matrix \(S_{jk} \in \{0,1\}\) encodes the initial rig of each machine:
\begin{align}
S_{jk} &=
\begin{cases}
1, & \text{if machine } j \text{ has initial rig } k,\\
0, & \text{otherwise}.
\end{cases}
\label{descr:Sjk}
\end{align} The resulting mean initial rig–change time on machine \(j\) is
\begin{equation}
  \bar S_j = \sum_{k=1}^L S_{jk} \,\bar R_k .
  \label{eq:model14_Sbar}
\end{equation}
Matrix \(r_{ik} \in \{0,1\}\) indicates the rig requirement of job \(i\):
\begin{align}
R_{ik} &=
\begin{cases}
1, & \text{if job } i \text{ is executed with rig configuration } k,\\
0, & \text{otherwise}.
\end{cases}
\label{descr:Sjk2}
\end{align} 
Objective 1 $|$ $H_{\mathrm{obj}}$: Minimise the deviation of the total duration
on each machine from the mean duration, including the initial rig configuration set on each machine
\begin{equation}
    H_{\mathrm{obj}}
    = \sum_{j=1}^M 
        \left(
            \sum_{i=1}^N x_{ij} d_i
            + \sum_{k=1}^L y_{kj} \bar{R}_k
            - \bar{S}_j
            - \bar{D}
        \right)^2.
\end{equation} 
Constraint 1 $|$ $H_{\mathrm{single}}$: Each job must be assigned to exactly one
machine
\begin{equation}
    H_{\mathrm{single}}
    = \sum_{i=1}^N 
        \left(
            \sum_{j=1}^M x_{ij} - 1
        \right)^2.
\end{equation}

Constraint 2 $|$ $H_{\mathrm{rig}}$: Which rig changes take place on which machine
\begin{equation}
    H_{\mathrm{rig}}
    = \sum_{j=1}^M 
        \sum_{k=1}^L 
            \left(
                S_{jk} + \sum_{i=1}^N R_{ik} x_{ij}
            \right)(1 - y_{kj}).
\end{equation}
Constraint 3 $|$ $H_{\mathrm{switch}}$: Minimise rig-change time between
the jobs on the same machine
\begin{equation}
    H_{\mathrm{pair}}
    = \sum_{j=1}^M
        \left(
            \sum_{k = 1}^L 
                \bar{R}_{k} y_{kj} - S_{j}
        \right)^2.
\end{equation}

\subsection{Classical Approximation Model}\label{subsec:model3}
For the classical implementation, we model the problem as a parallel-machine
scheduling problem with sequence-dependent setup times (PMSP-SDST) and
flexible job assignment. The formulation is based on established PMSP-SDST
literature~\cite{Ovacik1993,Kurz2001,Franca1996,Sivrikaya1999,Radhakrishnan2000,Bruni2020,YING2025100340}.
Our approach combines a greedy constructive heuristic, local improvement,
SDST-aware insertion, and an iterated greedy destroy-repair procedure.
First, jobs are grouped by rig and sorted within each rig by decreasing
processing time. Rig blocks are then assigned to machines using a score
that balances machine load and setup cost,
\begin{equation}
\lambda \cdot \text{load}(j) + \text{setup}(\text{tail\_rig}(j)\rightarrow\text{rig\_block}).
\end{equation}
The resulting machine sequences are improved by local two-swap search.
To further refine solutions, we apply an iterated greedy framework that
destroys and reconstructs partial schedules using SDST-aware insertion,
with occasional uphill acceptance based on
\begin{equation}
\text{Probability}(\text{accept}) = \exp(\Delta / T).
\label{eq:probability}
\end{equation}
For small instances, we additionally use a disjunctive
MILP baseline with sequence variables $y_{ijk}$,
dummy start jobs, and big-$M$ timing
constraints. The MILP minimises the makespan $C_{\max}$ subject to
standard sequencing constraints of the form
\begin{equation}
C_{\max} \ge S_i + p_i,
\qquad
S_k \ge S_i + p_i + s_{i\rightarrow k} - M(1-y_{ijk}).
\end{equation}

Algorithm~\ref{alg:ourmethod} summarises the overall heuristic framework.

\begin{algorithm}[t]
\caption{Heuristic Framework for PMSP-SDST Scheduling}
\label{alg:ourmethod}
\begin{algorithmic}[1]
\REQUIRE Instances with jobs, machines, processing times, and SDST matrix
\ENSURE A schedule minimising $C_{\max}$

\STATE Group jobs by rig; sort each rig in descending duration
\STATE Initialise empty sequence for each machine

\FOR{each rig block in random order}
    \FOR{each machine $j$ in parallel}
        \STATE Compute score $\lambda \cdot \mathrm{load}(j) + \mathrm{setup}(\mathrm{tail}(j)\rightarrow \mathrm{rig})$
    \ENDFOR
    \STATE Assign rig block to machine with minimum score
\ENDFOR

\FOR{each machine $j$}
    \STATE Perform 2-swap local search on the sequence of $j$
\ENDFOR

\WHILE{termination criterion not met}
    \STATE Destroy part of the schedule
    \FOR{each removed job}
        \STATE Evaluate all insertion positions using SDST marginal cost
        \STATE Insert at best position
    \ENDFOR
    \IF{new solution worse}
        \STATE Accept with probability $\exp(\Delta/T)$
    \ENDIF
\ENDWHILE

\STATE \textbf{return} best schedule found
\end{algorithmic}
\end{algorithm}

\subsection{QAOA}
QAOA is a hybrid variational algorithm for combinatorial optimisation on
NISQ hardware~\cite{Fahri_2014}. It alternates problem and mixer
Hamiltonians over $p$ layers, while a classical optimiser updates the
parameters to minimise the problem Hamiltonian expectation value. The
resulting output distribution yields an approximate solution to the QUBO
instance.
\section{Hardware Platforms} \label{sec:hardware_platforms}
\subsection{IBM Quantum (Eagle and Heron Processors)}
IBM’s gate-based devices use superconducting fixed-frequency transmon
qubits operated at millikelvin temperatures and arranged in a
heavy-hex topology to reduce crosstalk and improve
manufacturability~\cite{kjaergaard2020superconducting,krantz2019quantum,chamberland2020topological}.
Eagle provides 127 qubits, while Heron introduces redesigned couplers
and a new control stack for improved two-qubit
fidelities~\cite{ibm2023heron}.

\subsection{D-Wave Quantum Annealer (Advantage System)}
D-Wave’s \emph{Advantage} quantum annealer solves Ising/QUBO problems
through annealing dynamics and contains over 5000 superconducting qubits
in the Pegasus connectivity graph~\cite{johnson2011quantum,boothby2020next}.
It also offers a hybrid solver combining classical heuristics with
quantum annealing, although its internal optimisation process is
intransparent.

\subsection{Fujitsu Digital Annealer}
The Fujitsu Digital Annealer is a quantum-inspired CMOS accelerator that
emulates annealing dynamics at room temperature and performs parallel
updates on QUBO formulations~\cite{aramon2019physics}. Since it does not
require minor-embedding, it can directly represent dense optimisation
problems~\cite{fujitsuDAwhitepaper}.

\section{Experimental Setup}
To ensure comparability, we evaluate the proposed JSS formulation
using a unified experimental setup for quantum, quantum-inspired,
and classical solvers~\cite{Thelen:2025,Gierisch:2026}. This includes benchmark instances, solver
settings, parameters, validation metrics, and integration into
the classical industrial workflow~\cite{Schwenzow2024}.

\subsection{Benchmark}
All experiments use real industrial JSS instances provided by
Siemens. The instances range from small toy problems with $2$ machines,
$2$ rigs, and $2$ jobs to industry-scale and larger cases with $32$
machines, $85$ rigs, and $256$ jobs.
Each instance specifies machines, jobs, rigs, processing times,
and rig setup relations, with multiple instances per size to vary difficulty.
For small instances, a branch-and-bound solver provides the optimal makespan and
job-to-machine assignment. the largest tractable case has $6$ machines,
$5$ rigs, and $14$ jobs. For larger instances, we use the best solution
found within a one-hour time limit.
We run our experiments on the D-Wave Quantum Annealer, IBM Eagle and
IBM Heron processors, the Fujitsu Digital Annealer, and classical
hardware equipped with an Intel Xeon~E5{-}2683~v4 CPU (16~cores,
32~threads per socket; 2~sockets, 64~logical CPUs in total) running at
up to 3.0~GHz.
For the quantum and quantum-inspired evaluations, we implement two
assignment-based QUBO formulations, the Single-Constraint and
Multi-Constraint Models, described in
Sections~\ref{subsec:model-1} and~\ref{subsec:model-2}. Both encode only
the job-to-machine assignment and do not enforce internal job order. For
the considered instance sizes, sequencing is efficiently handled
classically~\cite{sorting_algorithms}, making this decomposition
suitable for a realistic hybrid industrial workflow. To ensure a fair
comparison with classical state-of-the-art methods, we also implement a
Classical Approximation Model (Section~\ref{subsec:model3}), which
approximates the assignment component efficiently on classical hardware.

\subsection{Hardware and Solver Setups}
\subsubsection{D-Wave Annealer}
The QUBO models are solved on the D-Wave Advantage and hybrid systems.
For each instance, we use $1024$ shots and vary the chain strength
according to D-Wave guidelines~\cite{dwave_chain_strength_2020}.

\subsubsection{IBM Gate-Based Quantum Processors}
QAOA circuits for the Single-Constraint Model are generated in Qiskit
$1.4$ and executed with $1024$ shots at different QAOA depths to
identify when noise from additional layers outweighs gains in solution
quality. Experiments are conducted on the Eagle processor \textit{Kiev}
and the Heron processors \textit{Fez} and \textit{Torino}.

\subsubsection{Fujitsu Digital Annealer}\label{subsub:FDA}
The third-generation Digital Annealer is used as a reference
quantum-inspired architecture. It may be viewed as a lower bound on
what future quantum annealers could achieve with improved hardware
stability, connectivity, and precision, while acknowledging important
architectural differences from physical quantum annealers in the threats
to validity section~\ref{threats}. Experiments use three QUBO
implementations: a \texttt{PyQUBO} model, a standard \texttt{BinPol}
formulation, and an extended \texttt{BinPol} version that separates
objective and constraint terms to exploit hardware-supported constraint
handling.

\subsubsection{Classical Approximation}
For larger-scale problems, we implement the classical approximation
algorithm described in Section~\ref{subsec:model3} under the same
assignment constraints as the Single-Constraint and Multi-Constraint
Models, using between 100 and 1200 iterations.

\subsection{Evaluation Metrics}
We use several metrics to compare performance across all platforms:
\begin{enumerate}
    \item \textbf{Share of best, valid, and invalid solutions}, where best refers to the best feasible solution found within the time limit,
    valid satisfies all constraints and invalid violate at least
    one constraint.
    \item \textbf{Optimality gap} relative to the best-known solution.
    \item \textbf{Execution time} (end-to-end execution time).
    \item \textbf{Scalability} with respect to problem and QUBO size.
\end{enumerate}

\section{Results}
We next discuss the results across the hardware platforms
described in Section~\ref{sec:hardware_platforms}.

\subsection{D-Wave Annealer Results}
\begin{figure*}[htbp]
\includegraphics{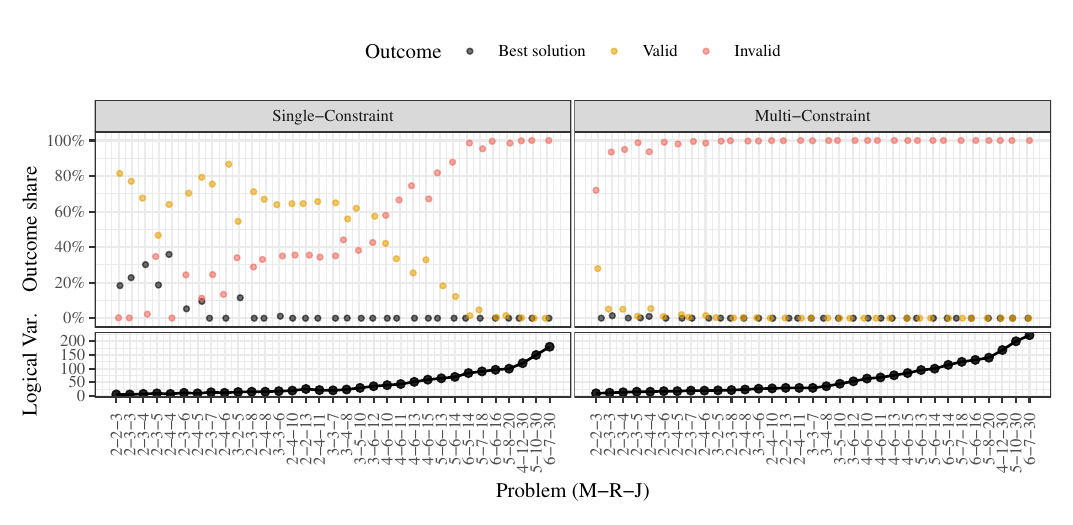}\vspace*{-3em}
    \caption{Scatter plot of best, valid, and invalid solution shares for the
    Single-Constraint and Multi-Constraint models on Advantage System 6.4.
    Problem instances on the x-axis are denoted as $(M\!-\!R\!-\!J)$ (machines-rigs-jobs).
    The plots below show the required logical qubits per model, with
    the Multi-Constraint model exhibiting higher qubit demand due to
    additional constraints. The y-axis represents the percentage of
    samples classified as best (black), valid (yellow), or invalid (red).}
    \label{fig:dwave_outcome_share}
\end{figure*}
Figure~\ref{fig:dwave_outcome_share} shows the performance of both QUBO
formulations on the D-Wave Advantage System 6.4. We also run
experiments on Advantage System 4.1, but the results are nearly
identical and are omitted for clarity. The $x$-axis lists problem
instances in increasing complexity as $(M, R, J)$, and the two plots
below show the required logical qubits. The
Multi-Constraint Model requires more qubits, scaling as $M \cdot J + M
\cdot R$, whereas the Single-Constraint Model uses only $M \cdot J$
variables. The $y$-axis reports the share of best, valid, and invalid
solutions. The Multi-Constraint Model is too constraint-dense for practical
execution on both Advantage generations: only very small toy instances
yield valid results, best solutions are rare, and invalid samples
quickly dominate as problem size increases. In contrast, the
Single-Constraint Model performs much better. Both Advantage systems 4.1
and 6.4 show nearly identical behaviour, indicating that the hardware
improvements between these generations are minimal for this problem class.
For the Single-Constraint Model, valid and best
solutions remain frequent up to QUBO sizes of about 100--120 qubits.

\begin{table}[t]
\centering
\scriptsize 
\setlength{\tabcolsep}{3pt}

\begin{minipage}{0.49\columnwidth}
\centering
\begin{tabular}{ll}
\hline
\textbf{Problem} & \textbf{Result} \\
\hline
M=2 R=2 J=3   & Best solution \\
M=2 R=3 J=3   & Best solution \\
M=2 R=3 J=4   & Best solution \\
M=2 R=3 J=5   & Best solution \\
M=2 R=4 J=4   & Best solution \\
M=2 R=3 J=6   & Valid \\
M=2 R=4 J=5   & Valid \\
M=2 R=3 J=7   & Valid \\
M=2 R=4 J=6   & Valid \\
M=3 R=2 J=5   & Best solution \\
M=2 R=3 J=8   & Valid \\
M=2 R=4 J=8   & Valid \\
M=3 R=3 J=6   & Valid \\
M=2 R=4 J=10  & Valid \\
M=2 R=2 J=13  & Valid \\
M=2 R=4 J=11  & Valid \\
\hline
\end{tabular}
\end{minipage}\hfill
\begin{minipage}{0.49\columnwidth}
\centering
\begin{tabular}{ll}
\hline
\textbf{Problem} & \textbf{Result} \\
\hline
M=3 R=3 J=7   & Valid \\
M=3 R=4 J=8   & Valid \\
M=3 R=5 J=10  & Valid \\
M=3 R=6 J=12  & Valid \\
M=4 R=6 J=10  & Valid \\
M=4 R=6 J=11  & Valid \\
M=4 R=6 J=13  & Valid \\
M=4 R=6 J=15  & Valid \\
M=5 R=6 J=14  & Valid \\
M=6 R=5 J=14  & Valid \\
M=5 R=7 J=18  & Valid \\
M=6 R=6 J=16  & Valid \\
M=5 R=8 J=20  & Valid \\
M=4 R=12 J=30 & Valid \\
M=5 R=10 J=30 & Valid \\
M=6 R=7 J=30  & Valid \\
\hline
\end{tabular}
\end{minipage}
\caption{Hybrid solver results for the Multi-Constraint model.}
\label{fig:dwave_hybrid_solver}
\end{table}

Table~\ref{fig:dwave_hybrid_solver} shows the D-Wave hybrid solver
results for the Multi-Constraint Model on the same instances. Across all
problem sizes, the solver returns at least one valid solution and, up to
about 15 qubits, the best solution. However, although 1024 shots are
executed, only the top solution is returned, so the share of invalid or
suboptimal results cannot be determined. Overall, the hybrid solver
produces valid schedules reliably across all tested instances.

\begin{figure*}[htbp]
\includegraphics{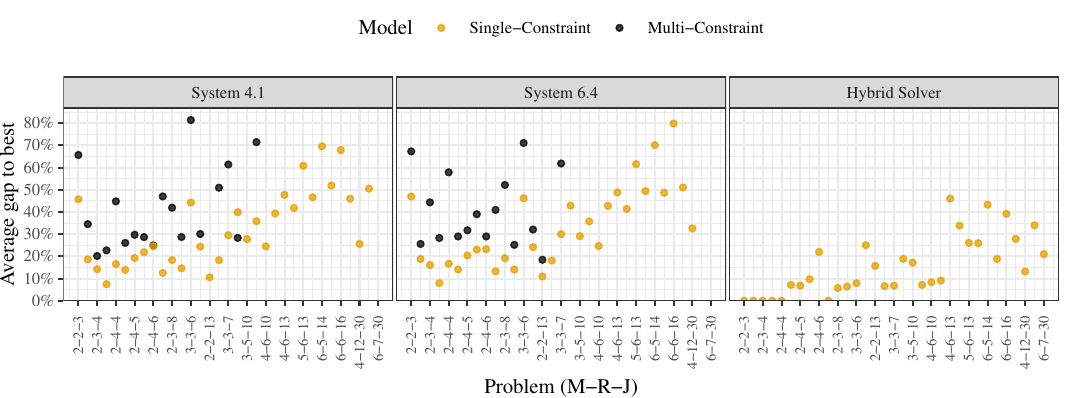}\vspace*{-2em}
    \caption{Average gap (in \%) 
    of all valid solutions relative
    to the best-known solution (y) for varying
    instances (x) in form $(M\!-\!R\!-\!J)$. Panels separate results
    for Advantage 4.1, Advantage 6.4, and the Hybrid Solver.
    The x-axis lists all problem instances; invalid
    and best solutions are excluded.}
    \label{fig:B4_gap_to_best_dwave}
\end{figure*}
Figure~\ref{fig:B4_gap_to_best_dwave} visualises the average
gap to the best-known solution, considering only valid solutions.
The goal of this analysis is to quantify, for each platform,
how close the valid solutions are to the optimal reference.
The two colours represent the two QUBO formulations.
Across both Advantage systems ($4.1$ and $6.4$), the Single-Constraint Model produces valid solutions closer to the best-known solution, mainly because the more faithful Multi-Constraint Model becomes too complex for D-Wave and yields too few valid solutions as problem size grows.
Later Fujitsu results show that this trend can reverse
when the hardware handles added complexity better.
The hybrid solver performs better than the pure quantum hardware:
its valid solutions exhibit smaller gaps to the best-known reference across
most instance sizes.
\begin{figure*}[htbp]
    \centering
    \includegraphics{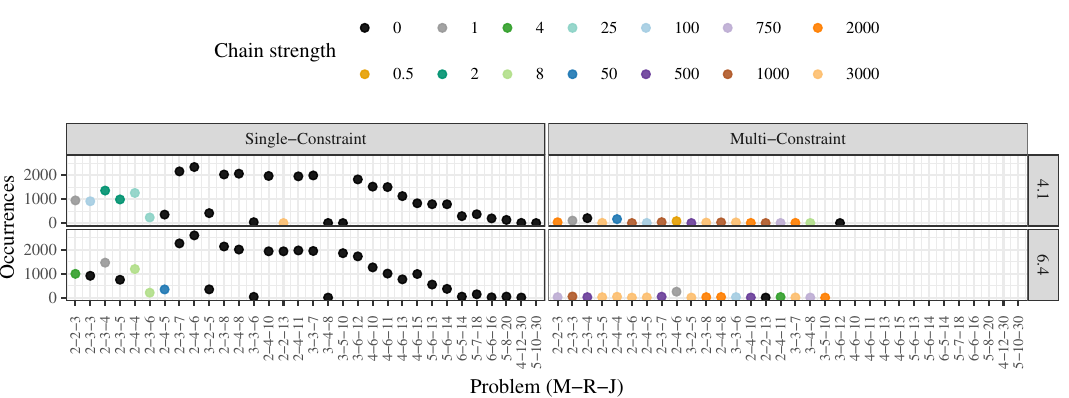}
    \caption{Impact of chain strength on the number of valid and best solutions for
    the Single-Constraint and Multi-Constraint models on Advantage 4.1 and
    6.4. The x-axis shows problem instances $(M\!-\!R\!-\!J)$, the y-axis
    the number of valid and best samples, and colours denote chain strength
    ($0$ = hardware default to $3000$).}
    \label{fig:winner_chain_strength}
\end{figure*}
Figure~\ref{fig:winner_chain_strength} compares the effect of different
chain strengths on both models across Advantage 4.1 and 6.4; chain
strength cannot be adjusted for the Hybrid Solver. For the
Multi-Constraint Model, chain strength influences solution quality,
whereas for the Single-Constraint Model the D-Wave default setting
(chain strength $0$) gives the most stable results. Manual tuning does
not yield systematic improvements and often reduces the number of valid
or best samples, especially for larger QUBOs. Overall, solvability
depends more on the problem formulation than on chain-strength tuning.

\subsection{IBM Gate-Based Quantum Computing}

\begin{figure}[htbp]
    \centering
    \includegraphics{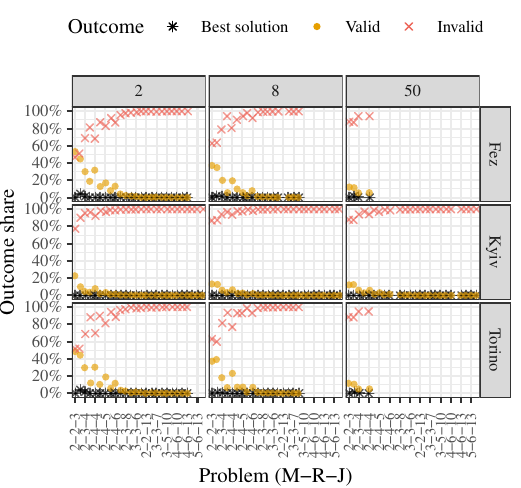}
    \caption{QAOA outcome shares on the Eagle (\textit{Kiev}) and Heron
    (\textit{Fez}, \textit{Torino}) processors. The x-axis shows problem
    instances \((M, R, J)\), the y-axis the percentage of best, valid, and
    invalid solutions, and each facet a different QAOA depth.}
    \label{fig:ibm_qaoa_outcomes}
\end{figure}
Figure~\ref{fig:ibm_qaoa_outcomes} summarises the results on the IBM
Eagle (\textit{Kiev}) and Heron (\textit{Fez}, \textit{Torino})
processors. The x-axis lists problem instances as \((M, R, J)\), the
y-axis shows the share of outcomes across all shots, and the facets
indicate the number of QAOA layers. We restrict the IBM experiments to
the Single-Constraint Model, as the Multi-Constraint Model does not
produce meaningful solutions on any tested system.
Across all processors, IBM hardware solves only small toy instances, and
the probability of obtaining valid or best solutions is only slightly
above random guessing. The Heron processors improve over Eagle, but not
enough to expand the range of solvable problems substantially.
Increasing the number of QAOA layers beyond a small threshold quickly
reduces performance due to noise~\cite{Greiwe:2023,Thelen:2024}. For problem sizes up to about $10$
qubits, up to $8$ layers may provide slight benefits,
but for larger problems additional layers introduce more noise than
improvement. Results for all tested layer
settings are available in the
\href{https://gitlab.oth-regensburg.de/IM/lfd/dissemination/phd/hila_safi/qc_in_industry_benchmark.git}{reproduction package}.

\subsection{Fujitsu Digital Annealer}
\begin{figure*}[htbp]
\includegraphics[width=\textwidth]{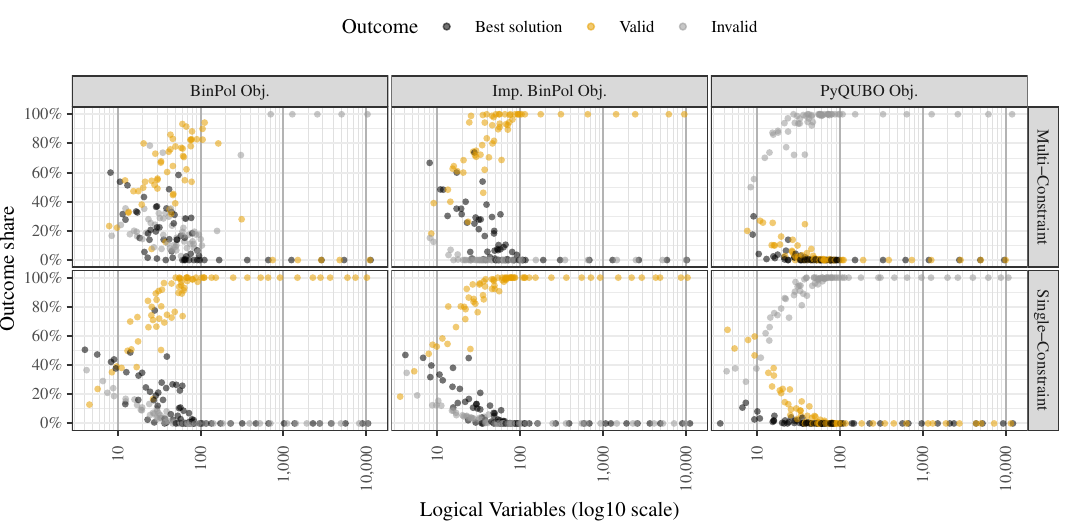}\vspace*{-1em}
    \caption{Overview of Fujitsu Digital Annealer results for the (\texttt{BinPol},
    improved \texttt{BinPol}, and \texttt{PyQUBO}) implementation
    and both problem formulations (Single-Constraint and Multi-Constraint).
    The $x$-axis shows the problem
    size (number of qubits) and the $y$-axis reports the share of outcomes
    classified as best, valid, or invalid.}
    \label{fig:fujitsu_overview}
\end{figure*}
Figure~\ref{fig:fujitsu_overview} summarises the results from the Fujitsu
Digital Annealer (DA) across the three QUBO implementations described in
Section~\ref{subsub:FDA}. Unlike the quantum devices, the DA supports
problem sizes beyond an equivalent of $10{,}000$ logical variables,
enabling large-scale experiments for both the Single-Constraint and
Multi-Constraint Models. The $x$-axis shows the number of logical
variables, and the $y$-axis the share of best, valid, and invalid
solutions.
For the Single-Constraint Model, both the standard \texttt{BinPol} and
the separated \texttt{BinPol} formulation perform consistently well
across all tested scales, returning high shares of valid and best
solutions even for the largest instances. For the more complex
Multi-Constraint Model, the standard \texttt{BinPol} formulation is no
longer sufficient at larger scales, and the separated \texttt{BinPol}
implementation is required to obtain valid solutions. By contrast, the
\texttt{PyQUBO} formulation degrades clearly for both models despite
being mathematically equivalent, indicating that internal compilation
and preprocessing in the DA software stack significantly affect solution
quality. As with the D-Wave hybrid solver, the DA returns only a limited
number of top solutions rather than the full sample distribution, but
the high rate of valid solutions across large instances shows that it is
well suited for the assignment-based formulations studied here.
\begin{figure}[htbp]    
\includegraphics{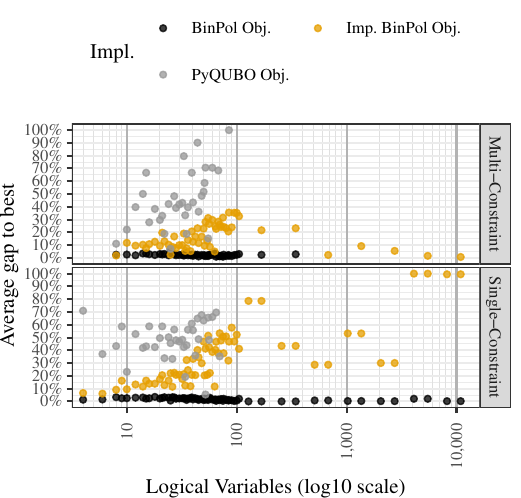}\vspace*{-1em}
    \caption{Average gap to the best-known solution for the Fujitsu
            Digital Annealer across problem sizes.}
    \label{fig:fujitsu_avg_gap}
\end{figure}
Figure~\ref{fig:fujitsu_avg_gap} shows the average gap of valid Fujitsu
Digital Annealer solutions to the best-known solution. The $x$-axis
gives the problem size in logical variables, the $y$-axis the average
relative deviation, colours denote the three QUBO implementations, and
the facets separate the Single-Constraint and Multi-Constraint Models.
Best and invalid solutions are excluded to focus on valid assignments.
Across both models, the \texttt{PyQUBO} formulation produces valid
solutions that are consistently farther from the best solution than the
two \texttt{BinPol}-based implementations. Although the Improved
\texttt{BinPol} variant performs best at finding optimal solutions,
especially for the Multi-Constraint Model, its average gap is larger. We
attribute this to the objective/constraint separation, which improves
feasibility but appears to reduce fine-grained optimisation
performance. The standard \texttt{BinPol} formulation achieves the
smallest average gap and thus the best valid-solution quality across a
wide range of problem sizes.
\begin{figure}[htbp]
\includegraphics{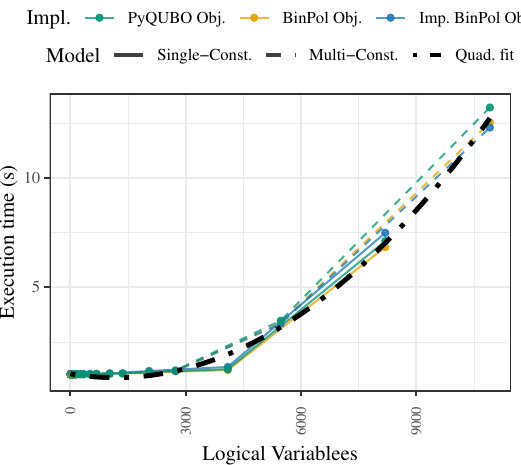}\vspace*{-1em}
    \caption{Execution time on the Fujitsu Digital Annealer as a
    function of problem size (number of logical variables).}
    \label{fig:p_exec_fujitsu}
\end{figure}
To assess the scaling behaviour of the Fujitsu Digital Annealer,
Figure~\ref{fig:p_exec_fujitsu} shows the execution time as a function
of QUBO size across all implementations and formulations. In all cases,
execution time exhibits a clear quadratic growth trend with the number
of logical variables.
\subsection{Classical Approximation}

\begin{figure*}[htbp]
\includegraphics{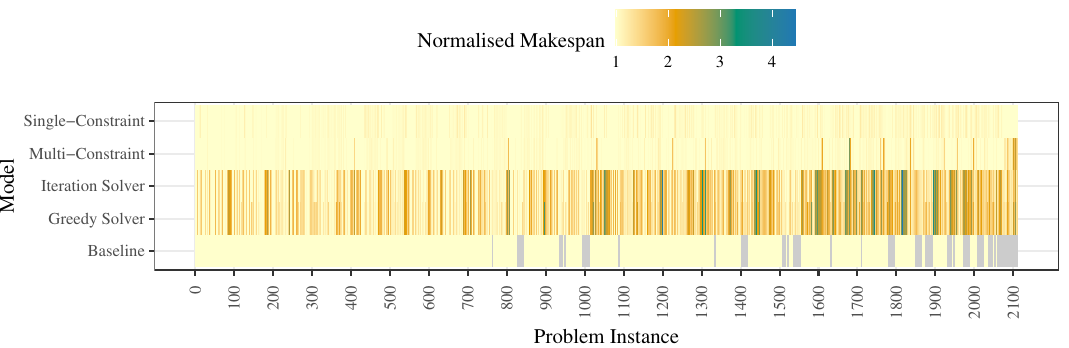}\vspace*{-1em}
    \caption{Comparison of normalised makespan across all problem instances for
    Fujitsu Digital Annealer (Single-Constraint and Multi-Constraint Model),
    the classical greedy and iterative solver, and the optimal solution as
    baseline. Lower values indicate
    better schedules.}
    \label{fig:comparison_fuj_classical}
\end{figure*}
Finally, we compare the classical approximation approach with the
Fujitsu Digital Annealer, as both solve industry-scale problems.
Figure~\ref{fig:comparison_fuj_classical} shows the normalised makespan
for the Single-Constraint and Multi-Constraint Models, the iterative
classical solver, the greedy solver, and the optimal reference as a
lower bound. Across nearly all instances, both Fujitsu-based models
outperform the classical approximation methods. This is particularly
relevant when considering runtime: The classical solvers also exhibit
quadratic empirical scaling, similar to the Fujitsu annealer. Even the
iterative version with 1000 iterations provides little improvement over
the greedy heuristic and remains consistently worse than the
annealer-based solutions. Overall, the Digital Annealer yields higher-quality
schedules for large-scale problems, when being evaluated under
the same runtime scaling setting.
\section{Threats to Validity}
These results should be interpreted with some care,
as their generalisability may be
limited~\cite{apel_threats, threats_to_validity}.
\subsection{Evaluation baseline}
In this study, we deliberately compare solution quality against
the true optimal solutions rather than the QUBO optima. This
represents a conservative assessment, as the QUBO formulation
itself may introduce approximation errors independent of the solver.
Our goal was to evaluate practical usefulness for real decision making
rather than performance relative to the encoded objective alone.
Consequently, the reported optimality gaps may overestimate solver
limitations but better reflect end-to-end application performance.
\subsection{Internal Validity}\label{threats}
One threat stems from our treatment of the Fujitsu Digital Annealer(DA) as a practical
lower bound for future fault-tolerant quantum annealing. While this assumption
is reasonable from one perspective, the DA's newest generation enables separation of
objective terms and penalty constraints, a capability that current physical
quantum annealers do not have. To mitigate this risk, we additionally
evaluated a full QUBO formulation on the DA.
Another concern relates to modelling choices. The JSSP admits multiple QUBO encodings,
and different weighting of penalty terms prioritise different industrial challenges.
Thus our conclusions may not generalise across all variants of
the JSSP or other scheduling problems.
Also, both the DA and the D-Wave Hybrid Solver return only the best solutions
rather than full sample distributions. While sufficient for optimisation evaluations,
this limits transparency and prevents a deeper analysis on solution quality.
\subsection{Construct Validity}
Comparisons across computation platforms inherently face challenges
because resource notions differ. While we try to ensure fairness by solving
identical industrial problems on each platform, certain formulations may naturally
benefit from one architecture more than another. Depending on what is
considered \enquote{fair}, this can be viewed as a threat to validity.
From an industrial standpoint, however, the size of solvable 
instances matters more than abstract resource counts.
Classical solver configurations introduce an additional threat:
The branch-and-bound solver is allowed
extended computation time to obtain ground-truth optimal solutions wherever
feasible. In contrast, the MILP solver
was deliberately time-limited to allow for comparison with quantum
and quantum-inspired solvers. While we see this as appropriate for our
objectives, this also implies MILP results do not necessarily represent
the best possible classical approximations that could be reached using
the same algorithm with more compute time.
Additionally, splitting the problem introduces a trade-off: while it reduces
complexity, it may prevent the global optimum from being reachable in certain cases.
We evaluate the distance to the best found solution to quantify the trade-off.

\subsection{External Validity}
The evaluation focuses on one industrial JSSP variant and many different
instances. Although the instances are highly realistic
and reflect constraints encountered in production environments, it remains a representative
rather than exhaustive dataset. Different industries---or even different production lines within
Siemens---may apply different constraint priorities or operate under different scheduling dynamics.
We solved a specialised problem using specific modelling, and the results demonstrate strong interactions
between model formulation and hardware behaviour. Although these trends may also apply to other optimisation
problems, as suggested by findings in previous work~\cite{safi2023},
they should still be interpreted with caution, and further verification across additional problem classes
is required.
\subsection{Ecological Validity}
Although we conceptually simulate the integration of quantum and
quantum-inspired solvers into existing
classical pipelines, we do not include communication latencies,
API overheads, or queueing delays in the
runtime measurements. In operational settings, these overheads may
influence end-to-end performance,
particularly for cloud-based quantum hardware.
The absence of such effects may therefore
overestimate real-world performance.
Finally, the data used in this study are realistic but not directly
sourced from a live production
system. Real deployments include additional complexities, such as
dynamic job arrivals.
\section{Discussion}\label{sec:Discussion}
Across all platforms, we observe a consistent co-design pattern:
more expressive and constraint-rich QUBO formulations capture the
industrial problem more accurately, but quickly become intractable on
current quantum hardware. For near-term industrial optimisation, the
central modelling question is therefore not how to encode the full
problem as accurately as possible, but which parts of the problem
should remain in the hardware-facing formulation and which should be
shifted to hybrid or classical processing stages.
This is most visible in the comparison between the Single-Constraint
and Multi-Constraint Models. The Multi-Constraint Model, with
\(M \cdot J + M \cdot R\) variables, places substantially higher
constraint pressure on the hardware and yields valid solutions only for
toy instances on D-Wave. By contrast, the Single-Constraint Model,
with \(M \cdot J\) variables, remains solvable across all platforms and
is therefore better suited to near-term hardware. A similar pattern
appears on the Fujitsu Digital Annealer: although it can process
problem sizes beyond $10{,}000$ logical variables, the more
constraint-dense formulation still requires a more carefully tailored
implementation. Overall, our results show that constraint pressure, at
least as much as raw problem size, limits scalability across all
evaluated platforms.
\subsection{Trade-Offs and Transferable Insights}
A central cross-platform result is the trade-off between solution
validity and proximity to the optimum. Simplified formulations increase
the probability of obtaining valid solutions, but can lead to larger
gaps to the best-known solution. More detailed formulations represent
the scheduling problem better, but current hardware often struggles to
navigate the resulting constrained landscape efficiently.
These findings suggest three transferable design principles for
near-term industrial quantum optimisation. First, compact formulations
that preserve the dominant optimisation structure while relaxing
secondary constraints are often more effective on current hardware than
more faithful but constraint-dense encodings. Second, solver
assessment should be performed end-to-end, including feasibility,
approximation quality, and practical execution effort, rather than
only on encoded objective values. Third, the most suitable solver
paradigm depends not only on problem size, but on the interaction
between constraint density, formulation structure, and hardware
capabilities. These observations are likely relevant beyond the present
JSSP case to a broader class of industrial assignment and scheduling
problems.
\subsection{Platform-Specific Lessons}
The IBM processors show architectural improvement from Eagle to Heron,
but for this problem class they still solve only toy instances, and
additional QAOA layers beyond small depths quickly become ineffective
because noise outweighs any gain in solution quality. D-Wave performs
substantially better with the compact formulation, while the
Multi-Constraint Model remains impractical beyond toy scales.
The Fujitsu Digital Annealer illustrates the benefit of specialised
hardware most clearly: with a suitable formulation, it handles
industry-scale instances and consistently outperforms the classical
approximation heuristics in solution quality.
From a practical perspective, the main lesson is not that one solver
paradigm universally dominates, but that successful industrial
proof-of-concept work currently depends on restricted problem scope,
hardware-aware formulation, and explicit matching of optimisation goals
to platform capabilities. Under present hardware conditions, this is
more important than abstract expectations of quantum advantage.
\subsection{Conclusion}
We conducted a multi-platform evaluation of quantum annealing,
quantum-inspired computing, and classical approximation methods for a
real-world JSSP variant. Our results show that current quantum hardware
is not yet competitive for large-scale constrained optimisation,
although simplified formulations allow non-trivial valid solutions.
Quantum-inspired hardware, by contrast, already supports
industry-relevant problem sizes for the assignment-based formulations
studied here and therefore provides a practically valuable reference
point for near-term industrial optimisation workflows. Overall, the
study highlights the importance of hardware-software co-design,
formulation tuning, and systematic design-space exploration.
\section{Outlook}
A next step is to move from case-specific studies towards systematic
design-space exploration across formulations, hardware backends, and
solver settings. Such work should separate formulation design, solver
configuration, and end-to-end integration effects more clearly, and
should account explicitly for whether the goal is optimal solutions,
high-quality approximations, or fast feasible schedules. Future work
should therefore extend the analysis to additional optimisation
problems, improved hybrid workflows, and live production settings,
including integration overheads and latency, to better understand the
practical value of quantum and quantum-inspired optimisation in
industrial environments.

This work was supported by the German Federal Ministry of Education and Research (BMBF), funding program ‘quantum technologies—from basic research to market’, grant numbers 13N16092 and 13N16093.

\bibliographystyle{IEEEtran}
\bibliography{bibcontrol, bibliography}

@misc{Fahri_2014,
 doi = {10.48550/ARXIV.1411.4028},
 url = {https://arxiv.org/abs/1411.4028},
 author = {Farhi, Edward and Goldstone, Jeffrey and Gutmann, Sam},
 title = {A Quantum Approximate Optimization Algorithm},
 publisher = {arXiv},
 year = {2014},
 copyright = {arXiv.org perpetual, non-exclusive license}
}

@inproceedings{safi2023,
 author = {Safi, Hila and Wintersperger, Karen and Mauerer, Wolfgang},
 booktitle = {2023 IEEE International Conference on Quantum Software (QSW)},
 doi = {10.1109/QSW59989.2023.00022},
 entrysubtype = {Conference},
 language = {en},
 pages = {104-115},
 title = {Influence of HW-SW-Co-Design on Quantum Computing Scalability},
 userd = {IEEE QSW '23},
 year = {2023},
 url = {https://www.lfdr.de/Publications/2023/SaWiMa23.pdf}
}

@article{YING2025100340,
 title = {Scheduling with sequence-dependent setup times in short-term production planning: A main path analysis-based review},
 journal = {Operations Research Perspectives},
 volume = {14},
 pages = {100340},
 year = {2025},
 issn = {2214-7160},
 doi = {https://doi.org/10.1016/j.orp.2025.100340},
 url = {https://www.sciencedirect.com/science/article/pii/S2214716025000168},
 author = {Kuo-Ching Ying and Pourya Pourhejazy and Zhi-Rong Lin}
}

@article{Ovacik1993,
 title = {Worst-case error bounds for parallel machine scheduling problems with bounded sequence-dependent setup times},
 journal = {Operations Research Letters},
 volume = {14},
 number = {5},
 pages = {251-256},
 year = {1993},
 issn = {0167-6377},
 doi = {https://doi.org/10.1016/0167-6377(93)90089-Y},
 url = {https://www.sciencedirect.com/science/article/pii/016763779390089Y},
 author = {Irfan M. Ovacik and Reha Uzhoy}
}

@article{Kurz2001,
 author = {M. E. Kurz and R. G. Askin},
 title = {Heuristic scheduling of parallel machines with sequence-dependent set-up times},
 journal = {International Journal of Production Research},
 volume = {39},
 number = {16},
 pages = {3747--3769},
 year = {2001},
 publisher = {Taylor \& Francis},
 doi = {10.1080/00207540110064938},
 URL = {https://doi.org/10.1080/00207540110064938},
 eprint = {https://doi.org/10.1080/00207540110064938}
}

@article{Franca1996,
 title = {A tabu search heuristic for the multiprocessor scheduling problem with sequence dependent setup times},
 journal = {International Journal of Production Economics},
 volume = {43},
 number = {2},
 pages = {79-89},
 year = {1996},
 issn = {0925-5273},
 doi = {https://doi.org/10.1016/0925-5273(96)00031-X},
 url = {https://www.sciencedirect.com/science/article/pii/092552739600031X},
 author = {Paulo M. França and Michel Gendreau and Gilbert Laporte and Felipe M. Müller}
}

@article{Sivrikaya1999,
 title = {Parallel machine scheduling with earliness and tardiness penalties},
 journal = {Computers \& Operations Research},
 volume = {26},
 number = {8},
 pages = {773-787},
 year = {1999},
 issn = {0305-0548},
 doi = {https://doi.org/10.1016/S0305-0548(98)00090-2},
 url = {https://www.sciencedirect.com/science/article/pii/S0305054898000902},
 author = {Funda Sivrikaya-Şerifoǧlu and Gündüz Ulusoy}
}

@article{Radhakrishnan2000,
 author = {Sanjay Radhakrishnan and Jose A. Ventura},
 title = {Simulated annealing for parallel machine scheduling with earliness-tardiness penalties and sequence-dependent set-up times},
 journal = {International Journal of Production Research},
 volume = {38},
 number = {10},
 pages = {2233--2252},
 year = {2000},
 publisher = {Taylor \& Francis},
 doi = {10.1080/00207540050028070},
 URL = {https://doi.org/10.1080/00207540050028070},
 eprint = {https://doi.org/10.1080/00207540050028070}
}

@article{Bruni2020,
 title = {The distributionally robust machine scheduling problem with job selection and sequence-dependent setup times},
 journal = {Computers \& Operations Research},
 volume = {123},
 pages = {105017},
 year = {2020},
 issn = {0305-0548},
 doi = {https://doi.org/10.1016/j.cor.2020.105017},
 url = {https://www.sciencedirect.com/science/article/pii/S0305054820301349},
 author = {M.E. Bruni and S. Khodaparasti and E. Demeulemeester}
}

@techreport{dwave_chain_strength_2020,
 title        = {Programming the {D}-{Wave} {QPU}: Setting the Chain Strength},
 institution  = {D-Wave Systems Inc.},
 author       = {D-Wave},
 address      = {Burnaby, BC, Canada},
 year         = {2020},
 month        = {April},
 url          = {https://www.dwavequantum.com/media/vsufwv1d/14-1041a-a_setting_the_chain_strength.pdf},
 note         = {White Paper}
}

@article{kjaergaard2020superconducting,
 title={Superconducting qubits: Current state of play},
 author={Kjaergaard, M. and Schwartz, M. E. and Braum{\"u}ller, J. and Krantz, P. and Wang, J. I.-J. and Gustavsson, S. and Oliver, W. D.},
 journal={Annual Review of Condensed Matter Physics},
 volume={11},
 pages={369--395},
 year={2020}
}

@article{krantz2019quantum,
 author = {Krantz, Philip and Kjaergaard, M. and Yan, F. and Orlando, T. and Gustavsson, Simon and Oliver, W.},
 year = {2019},
 month = {06},
 pages = {021318},
 title = {A quantum engineer's guide to superconducting qubits},
 volume = {6},
 journal = {Applied Physics Reviews},
 doi = {10.1063/1.5089550},
 url = {http://dx.doi.org/10.1063/1.5089550}
}

@article{chamberland2020topological,
 title = {Topological and Subsystem Codes on Low-Degree Graphs with Flag Qubits},
 author = {Chamberland, Christopher and Zhu, Guanyu and Yoder, Theodore J. and Hertzberg, Jared B. and  Cross, Andrew W.},
 journal = {Phys. Rev. X},
 volume = {10},
 issue = {1},
 pages = {011022},
 numpages = {19},
 year = {2020},
 month = {Jan},
 publisher = {American Physical Society},
 doi = {10.1103/PhysRevX.10.011022},
 url = {https://link.aps.org/doi/10.1103/PhysRevX.10.011022}
}

@misc{ibm2023heron,
 title={IBM Heron processor technology overview},
 author={IBM Quantum},
 year={2025},
 howpublished={\url{https://quantum.cloud.ibm.com/docs/de/guides/processor-types}},
 note={Accessed: 2025-07-01}
}

@article{johnson2011quantum,
 title={Quantum annealing with manufactured spins},
 author={Johnson, M. W. et al.},
 journal={Nature},
 volume={473},
 pages={194--198},
 year={2011}
}

@misc{boothby2020next,
 title={Next-Generation Topology of D-Wave Quantum Processors}, 
 author={Kelly Boothby and Paul Bunyk and Jack Raymond and Aidan Roy},
 year={2020},
 eprint={2003.00133},
 archivePrefix={arXiv},
 primaryClass={quant-ph},
 url={https://arxiv.org/abs/2003.00133}, 
}

@article{aramon2019physics,
 author = {Aramon, Maliheh and Rosenberg, Gili and Valiante, Elisabetta and Miyazawa, Toshiyuki and Tamura, Hirotaka and Katzgraber, Helmut},
 year = {2019},
 month = {04},
 pages = {48},
 title = {Physics-Inspired Optimization for Quadratic Unconstrained Problems Using a Digital Annealer},
 volume = {7},
 journal = {Frontiers in Physics},
 doi = {10.3389/fphy.2019.00048},
 url={http://dx.doi.org/10.3389/fphy.2019.00048}
}

@misc{fujitsuDAwhitepaper,
 title={Fujitsu Digital Annealer: Architecture and Applications},
 author={Fujitsu Laboratories},
 year={2020},
 howpublished={Fujitsu Technical Whitepaper}
}

@article{Arisha2001JobSS,
 author = {Muluk, Asmuliardi and Akpolat, Hasan and Xu, Jichao},
 year = {2003},
 month = {01},
 pages = {481-492},
 title = {Scheduling problems — An overview},
 volume = {12},
 journal = {Journal of Systems Science and Systems Engineering},
 doi = {10.1007/s11518-006-0149-z},
 url = {https://doi.org/10.1007/s11518-006-0149-z}
}

@inproceedings{schmidbauer:25:qce,
 author = {Lukas Schmidbauer and Carlos A. Riofrío and Florian Heinrich and Vanessa Junk and Ulrich Schwenk and Thomas Husslein and Wolfgang Mauerer},
 booktitle = {Proceedings of the IEEE International Conference on Quantum Computing and Engineering},
 entrysubtype = {Conference},
 eprint = {2504.16607},
 month = {5},
 title = {Path Matters: Industrial Data Meet Quantum Optimization},
 url = {https://arxiv.org/abs/2504.16607},
 userd = {IEEE QCE '25},
 year = {2025}
}

@inproceedings{schoenberger:25:pvldb,
 author = {Schönberger, Manuel and Trummer, Immanuel and Mauerer, Wolfgang},
 booktitle = {Proceedings of the VLDB Endowment},
 doi = {10.48550/arXiv.2510.20308},
 entrysubtype = {Conference},
 language = {en},
 month = {12},
 title = {Hybrid Mixed Integer Linear Programming for Large-Scale Join Order Optimisation},
 url = {https://arxiv.org/abs/2510.20308},
 userc = {CORE23:A*},
 userd = {VLDB '26},
 year = {2025}
}

@article{jssp_complexity,
 ISSN = {0364765X, 15265471},
 URL = {http://www.jstor.org/stable/3689278},
 author = {M. R. Garey and D. S. Johnson and Ravi Sethi},
 journal = {Mathematics of Operations Research},
 number = {2},
 pages = {117--129},
 publisher = {INFORMS},
 title = {The Complexity of Flowshop and Jobshop Scheduling},
 urldate = {2025-12-09},
 volume = {1},
 year = {1976}
}

@article{sorting_algorithms,
 author = {Khan, Mohsin and Shaheen, Samina and Qureshi, Furqan},
 year = {2014},
 month = {05},
 pages = {1-10},
 title = {Comparative Analysis of five Sorting Algorithms on the basis of Best Case, Average Case, and Worst Case},
 volume = {3},
 journal = {International Journal of Information Technology and Electrical Engineering 2306-708X},
 url= {https://api.semanticscholar.org/CorpusID:63715675}
}

@inproceedings{Sax:2020,
author = {Sax, Irmi and Feld, Sebastian and Zielinski, Sebastian and Gabor, Thomas and Linnhoff-Popien, Claudia and Mauerer, Wolfgang},
title = {Approximate approximation on a quantum annealer},
year = {2020},
isbn = {9781450379564},
publisher = {Association for Computing Machinery},
address = {New York, NY, USA},
url = {https://doi.org/10.1145/3387902.3392635},
doi = {10.1145/3387902.3392635},
booktitle = {Proceedings of the 17th ACM International Conference on Computing Frontiers},
pages = {108–117},
numpages = {10},
location = {Catania, Sicily, Italy},
series = {CF '20}
}

@inproceedings{threats_to_validity,
 author = {Teixeira, Eudis and Fonseca, Liliane and Soares, Sergio},
 title = {Threats to validity in controlled experiments in software engineering: what the experts say and why this is relevant},
 year = {2018},
 isbn = {9781450365031},
 publisher = {Association for Computing Machinery},
 address = {New York, NY, USA},
 url = {https://doi.org/10.1145/3266237.3266264},
 doi = {10.1145/3266237.3266264},
 booktitle = {Proceedings of the XXXII Brazilian Symposium on Software Engineering},
 pages = {52–61},
 numpages = {10},
 location = {Sao Carlos, Brazil},
 series = {SBES '18}
}

@inproceedings{apel_threats,
 author = {Wyrich, Marvin and Apel, Sven},
 title = {Evidence Tetris in the Pixelated World of Validity Threats},
 year = {2024},
 isbn = {9798400705670},
 publisher = {Association for Computing Machinery},
 address = {New York, NY, USA},
 url = {https://doi.org/10.1145/3643664.3648203},
 doi = {10.1145/3643664.3648203},
 booktitle = {Proceedings of the 1st IEEE/ACM International Workshop on Methodological Issues with Empirical Studies in Software Engineering},
 pages = {13–16},
 numpages = {4},
 location = {Lisbon, Portugal},
 series = {WSESE '24}
}

@misc{venturelli2016quantumannealingimplementationjobshop,
 title={Quantum Annealing Implementation of Job-Shop Scheduling}, 
 author={Davide Venturelli and Dominic J. J. Marchand and Galo Rojo},
 year={2016},
 eprint={1506.08479},
 archivePrefix={arXiv},
 primaryClass={quant-ph},
 url={https://arxiv.org/abs/1506.08479}, 
}

@article{SchwormWuGlattetal.2024,
 author    = {Philipp Schworm and Xiangqian Wu and Moritz Glatt and Jan C. Aurich},
 title     = {Solving flexible job shop scheduling problems in manufacturing with Quantum Annealing},
 journal   = {Production Engineering},
 number    = {17},
 pages     = {105 -- 115},
 publisher = {Springer Nature - Springer},
 doi       = {10.1007/s11740-022-01145-8},
 url  = {https://nbn-resolving.de/urn:nbn:de:hbz:386-kluedo-79093},
 year      = {2024},
}

@inproceedings{Orts_2022,
 author = {Orts, Francisco and Puertas, Antonio M. and Garz\'{o}n, Ester M. and Ortega, Gloria},
 title = {Quantum Annealing to Solve the Unrelated Parallel Machine Scheduling Problem},
 year = {2022},
 isbn = {978-3-031-30444-6},
 publisher = {Springer-Verlag},
 address = {Berlin, Heidelberg},
 url = {https://doi.org/10.1007/978-3-031-30445-3_14},
 doi = {10.1007/978-3-031-30445-3_14},
 booktitle = {Parallel Processing and Applied Mathematics: 14th International Conference, PPAM 2022, Gdansk, Poland, September 11–14, 2022, Revised Selected Papers, Part II},
 pages = {165–176},
 numpages = {12},
 location = {Gdansk, Poland}
}

@article{Rathore_25,
 title = {Load balancing for high performance computing using quantum annealing},
 author = {Rathore, Omer and Basden, Alastair and Chancellor, Nicholas and Kusumaatmaja, Halim},
 journal = {Phys. Rev. Res.},
 volume = {7},
 issue = {1},
 pages = {013067},
 numpages = {14},
 year = {2025},
 month = {01},
 publisher = {American Physical Society},
 doi = {10.1103/PhysRevResearch.7.013067},
 url = {https://link.aps.org/doi/10.1103/PhysRevResearch.7.013067}
}

@misc{Schwenzow2024,
 author = {Schwenzow, Tilmann and Wintersperger, Karen and Safi, Hila and Sicard, Oliver and Niedermeier, Christoph and Liebrecht, Christoph and Franke, Jörg and Reitelshöfer, Sebastian},
 year = {2024},
 month = {08},
 pages = {},
 journal = {unknown},
 title = {Investigating the Variational Quantum Eigensolver to solve scheduling for identical parallel machines with sequence-dependent setup-times},
 doi = {10.21203/rs.3.rs-4907806/v1}
}

@article{Kurowski_23,
 journal={European Journal of Operational Research},
 author={Krzysztof Kurowski and Tomasz Pecyna and Mateusz Slysz and Rafał Różycki and Grzegorz Waligóra and Jan Weglarz},
 title={Application of quantum approximate optimization algorithm to job shop scheduling problem},
 year={2023},
 month={None},
 pages={518-528},
 volume={310},
 number={2},
 doi={10.1016/j.ejor.2023.03.013},
 url={https://ideas.repec.org/a/eee/ejores/v310y2023i2p518-528.html},
}

@article{Feng07,
 author = {Lu, Feng and Marinescu, Dan C.},
 title = {An R || Cmax Quantum Scheduling Algorithm},
 year = {2007},
 issue_date = {Jun 2007},
 publisher = {Kluwer Academic Publishers},
 address = {USA},
 volume = {6},
 number = {3},
 issn = {1570-0755},
 url = {https://doi.org/10.1007/s11128-006-0048-8},
 doi = {10.1007/s11128-006-0048-8},
 journal = {Quantum Information Processing},
 month = jun,
 pages = {159–178},
 numpages = {20}
}

@Inbook{Mitchell2025,
 author="Mitchell, John E.",
 editor="Pardalos, Panos M.
 and Prokopyev, Oleg A.",
 title="Integer Programming: Branch-and-Cut Algorithms",
 bookTitle="Encyclopedia of Optimization",
 year="2025",
 publisher="Springer Nature Switzerland",
 address="Cham",
 pages="1--8",
 isbn="978-3-030-54621-2",
 doi="10.1007/978-3-030-54621-2_287-1",
 url="https://doi.org/10.1007/978-3-030-54621-2_287-1"
}

@misc{gurobi2026methods,
 author       = {{Gurobi Optimization, LLC}},
 title        = {Optimization Methods: Choose the Right Approach},
 year         = {2026},
 note         = {Accessed 2026-04-20},
 url          = {https://www.gurobi.com/resources/blog/optimization-methods-choosing-the-right-approach-with-gurobi}
}

@INPROCEEDINGS{Mauerer:2022,
  author={Mauerer, Wolfgang and Scherzinger, Stefanie},
  booktitle={IEEE SANER}, 
  title={1-2-3 Reproducibility for Quantum Software Experiments}, 
  year={2022},
  volume={},
  number={},
  pages={1247-1248},
  doi={10.1109/SANER53432.2022.00148}}

@inbook{Gabor:2019,
  title = {Assessing Solution Quality of 3SAT on a Quantum Annealing Platform},
  ISBN = {9783030140823},
  ISSN = {1611-3349},
  url = {http://dx.doi.org/10.1007/978-3-030-14082-3_3},
  DOI = {10.1007/978-3-030-14082-3_3},
  booktitle = {Quantum Technology and Optimization Problems},
  publisher = {Springer International Publishing},
  author = {Gabor,  Thomas and Zielinski,  Sebastian and Feld,  Sebastian and Roch,  Christoph and Seidel,  Christian and Neukart,  Florian and Galter,  Isabella and Mauerer,  Wolfgang and Linnhoff-Popien,  Claudia},
  year = {2019},
  pages = {23–35}
}

@INPROCEEDINGS{Greiwe:2023,
  author={Greiwe, Felix and Krüger, Tom and Mauerer, Wolfgang},
  booktitle={IEEE QSW}, 
  title={Effects of Imperfections on Quantum Algorithms: A Software Engineering Perspective}, 
  year={2023},
  volume={},
  number={},
  pages={31-42},
  doi={10.1109/QSW59989.2023.00014}
}

@inbook{Yue:2023,
  title = {Challenges and Opportunities in Quantum Software Architecture},
  ISBN = {9783031368479},
  url = {http://dx.doi.org/10.1007/978-3-031-36847-9_1},
  DOI = {10.1007/978-3-031-36847-9_1},
  booktitle = {Software Architecture},
  publisher = {Springer Nature Switzerland},
  author = {Yue,  Tao and Mauerer,  Wolfgang and Ali,  Shaukat and Taibi,  Davide},
  year = {2023},
  pages = {1–23}
}

@INPROCEEDINGS{Thelen:2024,
  author={Thelen, Simon and Safi, Hila and Mauerer, Wolfgang},
  booktitle={IEEE QCE}, 
  title={Approximating under the Influence of Quantum Noise and Compute Power}, 
  year={2024},
  volume={02},
  number={},
  pages={274-279},
  doi={10.1109/QCE60285.2024.10291}
}

@article{blekos_review_2024,
	title = {A review on {Quantum} {Approximate} {Optimization} {Algorithm} and its variants},
	volume = {1068},
	issn = {0370-1573},
	url = {https://www.sciencedirect.com/science/article/pii/S0370157324001078},
	doi = {10.1016/j.physrep.2024.03.002},
	urldate = {2024-06-16},
	journal = {Physics Reports},
	author = {Blekos, Kostas and Brand, Dean and Ceschini, Andrea and Chou, Chiao-Hui and Li, Rui-Hao and Pandya, Komal and Summer, Alessandro},
	month = jun,
	year = {2024},
	pages = {1--66},
}

@article{zhou2020quantum,
  title = {Quantum Approximate Optimization Algorithm: Performance,  Mechanism,  and Implementation on Near-Term Devices},
  volume = {10},
  ISSN = {2160-3308},
  url = {http://dx.doi.org/10.1103/physrevx.10.021067},
  DOI = {10.1103/physrevx.10.021067},
  number = {2},
  journal = {Physical Review X},
  publisher = {American Physical Society (APS)},
  author = {Zhou,  Leo and Wang,  Sheng-Tao and Choi,  Soonwon and Pichler,  Hannes and Lukin,  Mikhail D.},
  year = {2020},
  month = jun
}

@article{Zhang:2017,
  title = {Near-optimal quantum circuit for Grover's unstructured search using a transverse field},
  author = {Jiang, Zhang and Rieffel, Eleanor G. and Wang, Zhihui},
  journal = {Phys.\ Rev.\ A},
  volume = {95},
  issue = {6},
  numpages = {8},
  year = {2017},
  month = {Jun},
  publisher = {American Physical Society},
  doi = {10.1103/PhysRevA.95.062317},
  url = {https://link.aps.org/doi/10.1103/PhysRevA.95.062317}
}

@article{Wang:2020,
  title = {XY mixers: Analytical and numerical results for the quantum alternating operator ansatz},
  volume = {101},
  ISSN = {2469-9934},
  url = {http://dx.doi.org/10.1103/PhysRevA.101.012320},
  DOI = {10.1103/physreva.101.012320},
  number = {1},
  journal = {Physical Review A},
  publisher = {American Physical Society (APS)},
  author = {Wang,  Zhihui and Rubin,  Nicholas C. and Dominy,  Jason M. and Rieffel,  Eleanor G.},
  year = {2020},
  month = jan 
}

@INPROCEEDINGS{Baertschi:2020,
  author={Bärtschi, Andreas and Eidenbenz, Stephan},
  booktitle={2020 IEEE International Conference on Quantum Computing and Engineering (QCE)}, 
  title={Grover Mixers for QAOA: Shifting Complexity from Mixer Design to State Preparation}, 
  year={2020},
  volume={},
  number={},
  pages={72-82},
  doi={10.1109/QCE49297.2020.00020}
}

@inproceedings{bravyi2020obstacles,
  title={Obstacles to variational quantum optimization from symmetry protection},
  author={Bravyi, Sergey and Kliesch, Alexander and Koenig, Robert and Tang, Eugene},
  booktitle={Physical Review Letters},
  volume={125},
  number={26},
  pages={260505},
  year={2020},
  organization={APS}
}

@article{Egger:2021,
  author={Egger, Daniel J. and Mareček, Jakub and Woerner, Stefan},
  journal={Quantum}, 
  title={Warm-starting quantum optimization}, 
  year={2021},
  volume={5},
  number={479},
  pages={479},
  doi={10.22331/q-2021-06-17-479}
}

@article{Awasthi:2023,
  title = {Quantum walks with iterated open quantum dynamics},
  volume = {7},
  ISSN = {2521-327X},
  url = {http://dx.doi.org/10.1038/s42005-023-01453-0},
  DOI = {10.1038/s42005-023-01453-0},
  number = {1},
  journal = {Communications Physics},
  publisher = {Springer Science and Business Media LLC},
  author = {Awasthi,  Abhinav and Bärtschi,  Andreas and Eidenbenz,  Stephan and Le,  Eric and Zhu,  Shuchen},
  year = {2023},
  month = dec 
}

@article{Tate:2023,
  title = {Bridging classical and quantum with SDP initialized warm-starts for QAOA},
  volume = {7},
  ISSN = {2521-327X},
  url = {http://dx.doi.org/10.1038/s42005-023-01439-y},
  DOI = {10.1038/s42005-023-01439-y},
  number = {1},
  journal = {Communications Physics},
  publisher = {Springer Science and Business Media LLC},
  author = {Tate,  Reuben and Moondra,  Majad and Gard,  Bryan and Mohseni,  M. and Economou,  Sophia E. and Barnes,  Edwin},
  year = {2023},
  month = nov 
}

@misc{Lorenz:2025,
      title={Systematic benchmarking of quantum computers: status and recommendations}, 
      author={Jeanette Miriam Lorenz and Thomas Monz and Jens Eisert and Daniel Reitzner and Félicien Schopfer and Frédéric Barbaresco and Krzysztof Kurowski and Ward van der Schoot and Thomas Strohm and Jean Senellart and Cécile M. Perrault and Martin Knufinke and Ziyad Amodjee and Mattia Giardini},
      year={2025},
      eprint={2503.04905},
      archivePrefix={arXiv},
      primaryClass={quant-ph},
      url={https://arxiv.org/abs/2503.04905}, 
}

@misc{Vijendran:2024,
  doi = {10.48550/ARXIV.2408.02342},
  url = {https://arxiv.org/abs/2408.02342},
  author = {Vijendran,  Vishnu and Das,  Riddhi S. and Hollenberg,  Lloyd C. L. and Hill,  Charles D. and Thompson,  Jayne},
  title = {Large Scale Quantum Approximate Optimization Algorithm on Non-planar Graphs with Machine Learning Noise Mitigation},
  publisher = {arXiv},
  year = {2024},
  copyright = {Creative Commons Attribution 4.0 International}
}

@misc{Sud:2024,
  title={QAOA-in-QAOA: solving large-scale MaxCut problems on small quantum machines}, 
  author={Jay Sud and Matthew P. Harrigan and Nicholas Rubin and Norm M. Tubman and Daniel Lidar and Emanuel Knill and Ryan Babbush},
  year={2024},
  eprint={2405.05087},
  archivePrefix={arXiv},
  primaryClass={quant-ph},
  url={https://arxiv.org/abs/2405.05087}
}

@misc{montanez2024towards,
  title={Towards a Universal QAOA Protocol: Evidence of quantum advantage in solving combinatorial optimization problems}, 
  author={Berlain Montanez-Barrera and Kristina Kottmann and Román Orús and Victor Dunjko and Marta Mauri},
  year={2024},
  eprint={2405.09169},
  archivePrefix={arXiv},
  primaryClass={quant-ph},
  url={https://arxiv.org/abs/2405.09169}
}

@misc{Fingar:2024,
  doi = {10.48550/ARXIV.2406.20084},
  url = {https://arxiv.org/abs/2406.20084},
  author = {Fingar,  Constantin and Müller,  Timon and Zanger,  Benedikt and Lechner,  Wolfgang},
  title = {A Parity Quantum Approximate Optimization Algorithm},
  publisher = {arXiv},
  year = {2024},
  copyright = {Creative Commons Attribution 4.0 International}
}

@article{streif2020training,
  title={Training the quantum approximate optimization algorithm without access to a quantum processing unit},
  author={Streif, Michael and Leib, Martin},
  journal={Quantum Science and Technology},
  volume={5},
  number={3},
  pages={034008},
  year={2020},
  publisher={IOP Publishing}
}

@article{bayerstadler:2021,
  title = {Industry quantum computing applications},
  volume = {8},
  ISSN = {2196-0763},
  url = {http://dx.doi.org/10.1140/epjqt/s40507-021-00114-x},
  DOI = {10.1140/epjqt/s40507-021-00114-x},
  number = {1},
  journal = {EPJ Quantum Technology},
  publisher = {Springer Science and Business Media LLC},
  author = { and Bayerstadler,  Andreas and Becquin,  Guillaume and Binder,  Julia and Botter,  Thierry and Ehm,  Hans and Ehmer,  Thomas and Erdmann,  Marvin and Gaus,  Norbert and Harbach,  Philipp and Hess,  Maximilian and Klepsch,  Johannes and Leib,  Martin and Luber,  Sebastian and Luckow,  Andre and Mansky,  Maximilian and Mauerer,  Wolfgang and Neukart,  Florian and Niedermeier,  Christoph and Palackal,  Lilly and Pfeiffer,  Ruben and Polenz,  Carsten and Sepulveda,  Johanna and Sievers,  Tammo and Standen,  Brian and Streif,  Michael and Strohm,  Thomas and Utschig-Utschig,  Clemens and Volz,  Daniel and Weiss,  Horst and Winter,  Fabian},
  year = {2021},
  month = Nov 
}

@article{kruger:2025,
  author        = {Krüger, Tom and Mauerer, Wolfgang},
  title         = {{Out of the Loop: Structural Approximation of Optimisation
                    Landscapes and non-Iterative Quantum Optimisation}},
  journal       = {Quantum},
  volume        = {9},
  pages         = {1903},
  month         = nov,
  year          = {2025},
  issn          = {2521-327X},
  publisher     = {Verein zur F{\"o}rderung des Open Access Publizierens
                   in den Quantenwissenschaften},
  doi           = {10.22331/q-2025-11-06-1903},
  url           = {https://doi.org/10.22331/q-2025-11-06-1903},
  eprint        = {2408.06493},
  archiveprefix = {arXiv},
  primaryclass  = {quant-ph}
}

@inproceedings{Krueger:2020,
  author    = {Krüger, Tom and Mauerer, Wolfgang},
  title     = {{Quantum Annealing-Based Software Components:
                An Experimental Case Study with SAT Solving}},
  booktitle = {Proceedings of the IEEE/ACM 42nd International Conference
               on Software Engineering Workshops},
  series    = {ICSEW'20},
  pages     = {445--450},
  numpages  = {6},
  location  = {Seoul, Republic of Korea},
  address   = {New York, NY, USA},
  publisher = {Association for Computing Machinery},
  year      = {2020},
  isbn      = {9781450379632},
  doi       = {10.1145/3387940.3391472},
  url       = {https://doi.org/10.1145/3387940.3391472}
}

@article{Schmidbauer:2026,
  author        = {Schmidbauer, Lukas and Lobe, Elisabeth and
                   Schaefer, Ina and Mauerer, Wolfgang},
  title         = {{It's Quick to be Square:
                    Fast Quadratisation for Quantum Toolchains}},
  journal       = {ACM Transactions on Quantum Computing},
  volume        = {7},
  number        = {3},
  articleno     = {18},
  pages         = {1--46},
  numpages      = {46},
  address       = {New York, NY, USA},
  publisher     = {Association for Computing Machinery},
  year          = {2026},
  doi           = {10.1145/3800943},
  url           = {https://doi.org/10.1145/3800943},
  eprint        = {2411.19934},
  archiveprefix = {arXiv},
  primaryclass  = {quant-ph}
}

@inproceedings{Schmidbauer:2025,
  author        = {Schmidbauer, Lukas and Mauerer, Wolfgang},
  title         = {{SAT Strikes Back:
                    Parameter and Path Relations in Quantum Toolchains}},
  booktitle     = {2025 IEEE International Conference on Quantum Software
                   (QSW)},
  pages         = {1--12},
  address       = {Los Alamitos, CA, USA},
  publisher     = {IEEE Computer Society},
  month         = jul,
  year          = {2025},
  doi           = {10.1109/QSW67625.2025.00021},
  url           = {https://doi.org/10.1109/QSW67625.2025.00021},
  eprint        = {2505.22060},
  archiveprefix = {arXiv},
  primaryclass  = {quant-ph}
}

@inproceedings{Gierisch:2026,
  author        = {Gierisch, Vincent and Mauerer, Wolfgang},
  title         = {{QEF: Reproducible and Exploratory Quantum Software
                    Experiments}},
  booktitle     = {Service-Oriented Computing -- ICSOC 2025 Workshops},
  series        = {Lecture Notes in Computer Science},
  volume        = {16441},
  publisher     = {Springer Singapore},
  year          = {2026},
  isbn          = {978-981-92-1766-3},
  note          = {Forthcoming},
  eprint        = {2511.04563},
  archiveprefix = {arXiv},
  primaryclass  = {quant-ph},
  url           = {https://arxiv.org/abs/2511.04563}
}

@inproceedings{Thelen:2025,
  author        = {Thelen, Simon and Mauerer, Wolfgang},
  title         = {{Predict and Conquer:
                    Navigating Algorithm Trade-Offs with
                    Quantum Design Automation}},
  booktitle     = {2025 IEEE International Conference on Quantum Computing
                   and Engineering (QCE)},
  volume        = {1},
  pages         = {591--602},
  address       = {Los Alamitos, CA, USA},
  publisher     = {IEEE Computer Society},
  year          = {2025},
  doi           = {10.1109/QCE65121.2025.00071},
  url           = {https://doi.org/10.1109/QCE65121.2025.00071},
  eprint        = {2507.06758},
  archiveprefix = {arXiv},
  primaryclass  = {quant-ph}
}

\end{document}